\documentclass{iopart}

\pdfoutput=1

\usepackage{color}
\usepackage{graphicx,epsfig}
\usepackage{bm}
\usepackage{amssymb}

\def\be{\begin{equation}}
\def\ee{\end{equation}}

\def\bea{\begin{eqnarray}}
\def\eea{\end{eqnarray}}

\def\a{\alpha}
\def\e{\epsilon}

\def\nn{\nonumber\\}

\begin{document}
\title[Corrections to scaling for block entanglement in massive spin-chains]
{Corrections to scaling for block entanglement in massive spin-chains}
\author{Pasquale Calabrese$^1$, John Cardy$^2$ and Ingo Peschel$^3$}
\address{$^1$Dipartimento di Fisica dell'Universit\`a di Pisa
and INFN, Pisa, Italy

   $^2$Oxford University, Rudolf Peierls Centre for
      Theoretical Physics, 1 Keble Road, Oxford, OX1 3NP, United Kingdom and All Souls College, Oxford.

$^3$Fachbereich Physik, Freie Universit\"at Berlin, Arnimallee 14, D-14195 Berlin, Germany}

\date{\today}

\begin{abstract}
We consider the R\'enyi entropies $S_n$ in one-dimensional massive
integrable models diagonalizable by means of corner transfer matrices (as Heisenberg and Ising
spin chains).
By means of explicit examples and using the relation of corner transfer matrix with the Virasoro algebra,
we show that close to a conformal invariant critical point, when the correlation length $\xi$ is finite but large,
the corrections to the scaling are of the {\it unusual} form $\xi^{-x/n}$, with $x$ the dimension of a relevant
operator in the conformal theory.
This is reminiscent of the results for gapless chains and should be valid for any massive
one-dimensional model close to a conformal critical point.
\end{abstract}

\maketitle
\section{Introduction}

In the last decade there has been an increasing interest in quantifying the amount of entanglement
that is present in the ground state of an extended quantum system \cite{Rev}.
A measure of the bipartite entanglement is given by the so-called R\'enyi entropies, defined as follows.
Let $|\Psi\rangle$ be the ground state of an extended quantum mechanical
system and $\rho=|\Psi\rangle\langle\Psi|$ its density matrix.
One then divides the Hilbert space into a part ${\cal A}$ and its complement
${\cal B}$ and considers the reduced density matrix $\rho_{\cal A}={\rm Tr}_{\cal B}\,\rho$ of
subsystem ${\cal A}$.
Finally, the R\'enyi entropies are given by
\be
S_n=\frac1{1-n}\ln{\rm Tr}\,\rho_{\cal A}^n\, .
\label{Sndef}
\ee
The particular case $n=1$ of (\ref{Sndef}) is the von Neumann
entropy $S_1$ and it is usually called simply {\it entanglement entropy}.
However, the knowledge of $S_n$ for different $n$ characterizes the
full spectrum of non-zero eigenvalues of $\rho_{\cal A}$ (see e.g.
\cite{cl-08}) and provides significantly more information than $S_1$.

The case that has been most widely studied is a critical one-dimensional system whose
continuum limit is described by a conformal field theory (CFT) of central charge $c$.
When ${\cal A}$ is an interval of length $\ell$
embedded in an infinite system, the asymptotic large-$\ell$ behavior
of the R\'enyi entropies is given by \cite{Holzhey,cc-04,cc-rev}
\begin{equation}
\label{Renyi:asymp}
S_n(\ell)
\simeq \frac{c}6\left(1+\frac1n \right)\ln \ell+c'_n\,,
\end{equation}
where $c_n'$ is a non-universal constant. This behavior has been verified analytically and numerically
in numerous models. Only more recently it has been realized that such asymptotic behavior can be
difficult to observe because of the presence
of large and {\it unusual} corrections to the scaling behaving like \cite{ccen-10,cc-10,ce-10}
\begin{equation}
S_n(\ell)\simeq \frac{c}6\left(1+\frac1n \right)\ln \ell+c'_n+ b_n \ell^{-2x/n}\,,
\label{Sncorr}
\end{equation}
where $x$ is the scaling dimension of a relevant operator (i.e. $x<2$) and $b_n$ another
non-universal constant. These power-law corrections decay slowly even for moderate
values of $n$ and completely obscure the asymptotic result for large $n$.
In field theory language, this unusual scaling originates from the conical singularities (at the boundary
between ${\cal A}$ and ${\cal B}$) present in the $n$-sheeted Riemann surface describing
${\rm Tr} \rho_{\cal A}^n$ \cite{cc-10}.

However, universal scaling is not a prerogative of the gapless models.
A nearby critical point  influences a part of the parameter space called `critical region' in which the
correlation length $\xi$ (inverse gap) is large but finite.
Simple scaling arguments suggest that when an infinite system is divided
in two semi-infinite halves, the entanglement entropy should scale as
\begin{equation}
\label{Renyi:asymp-gap}
S_n \simeq \frac{c}{12}\left(1+\frac1n \right)\ln \xi+C'_n\,,
\end{equation}
where $C'_n$ is yet another non-universal constant. (One could also fix the normalization of the correlation
length in such a way that $C'_n=c'_n/2$ as done in some practical instances \cite{cv-10}.)
This formula has indeed been corroborated by a general field-theory argument that parallels
the c-theorem \cite{cc-04} and it has been verified for many integrable models.
In this massive case, the spatial division in  ${\cal A}$ and ${\cal B}$ is not as important as in the
massless case.
Indeed as long as the correlation length is smaller than all the separations (e.g. $\ell$ above), the
R\'enyi entropies of many disjoint blocks are given by Eq. (\ref{Renyi:asymp-gap}) multiplied by
the number of boundary points between ${\cal A}$ and ${\cal B}$. This is nothing but the one-dimensional
area law.
The question that naturally arises is whether {\it unusual} corrections to the scaling
in $\xi$ would be also present for the massive case. By simple scaling hypothesis, one could
argue that corrections due to a finite correlation length in Eq. (\ref{Sncorr}) should be functions
of $\xi/\ell$ and then changing the limit from $\ell\ll \xi$ to $\ell \gg \xi$, one argues
\begin{equation}
\label{Sgapcorr}
S_n\simeq \frac{c}{12}\left(1+\frac1n \right)\ln \xi+C'_n+B_n \xi^{-x/n}\,.
\end{equation}
(Notice that the exponent of the corrections is half the one in (\ref{Sncorr}) because of the presence
of a single conical singularity, analogously to the case of critical systems with boundaries \cite{cc-10}).
However, because of the very unusual form of these corrections, one could  doubt whether such
a simple scaling argument gives the correct answer.\footnote{This scaling analysis is corroborated by recent results for gapless systems in a trapping
potential \cite{cv-10}.
In this case, the actual system is gapless, but a length scale is generated by the trapping potential
and the resulting scaling is exactly given by Eq. (\ref{Sgapcorr}) where $\xi$ is replaced by the
trap-scale.}
Here we provide the analytical evidence that in some integrable models in which
$S_n$ can be obtained through the Baxter corner transfer matrix \cite{Baxter} this scaling is indeed
correct and that further corrections are of the form $\xi^{-kx/n}$ and $\xi^{-kx}$ with $k$ integer.

\section{Corrections to the scaling of R\'enyi entropies using corner transfer matrix}

When dealing with the geometric bipartition considered in this paper (i.e. two semi-infinite half lines)
the corner transfer matrix (CTM)-- that is a classical tool of statistical mechanics \cite{Baxter}-- helps
considerably in the derivation of the reduced density matrix $\rho_{\cal A}$ and hence for the
characterization of the entanglement entropies \cite{cc-04,pe-rev}.

In two-dimensional statistical models the CTM $\hat A$ is the
transfer matrix that connects an horizontal row (let say the line
$x<0$)  to a vertical one (let say $y<0$). Rigorously speaking,
each element of $\hat A$ is the partition function of the
left-lower quadrant when the spins on the negative $x$ and $y$
axes are fixed to given values. If, following Baxter
\cite{Baxter}, we choose the lattice in a clever way (i.e. rotated
by $\pi/4$ with respect to the axes), the four corner transfer
matrices are all equivalent (under some symmetry requirements on
the model) and the full partition function is just ${\rm Tr} {\hat
A}^4$. It is also evident that the reduced density matrix of the
quantum problem whose time-discretized version is the classical
model under consideration (see e.g \cite{nishino,pkl-99}) is
\be\label{eqrhoA}
\rho_{\cal A}= \frac{{\hat A}^4}{{\rm Tr}{\hat
A}^4}\,,
\ee
where the denominator is fixed by the normalization
${\rm Tr}\rho_{\cal A}=1$.

For several models, it is possible to obtain this reduced density
matrix exactly in the infinite-lattice limit..
It always  assumes the form \cite{pkl-99,pe-rev}
\be
\rho_{\cal A}=\frac{e^{-H_{\rm CTM}}}{{\rm Tr}\, e^{-H_{\rm CTM}}}\,,
\ee
where ${\hat H}_{\rm CTM}$ is an effective Hamiltonian, which, for many interesting integrable massive
models, may be rewritten in terms of free-fermionic operators (see \cite{pkl-99} and references therein)
\be
H_{\rm CTM}=\sum_{j=0}^\infty \e_j n_j\,.
\ee
where $n_j=c_i^\dagger c_i$ are  fermion occupation numbers  and have eigenvalues $0$ and $1$.
These properties are very general features of a large class
of massive integrable spin-chains.
The specificity of the model comes from the functional form of the ``single-particle levels'' $\e_j$.
In this basis $\rho_{\cal A}$ is diagonal and so the R\'enyi entropies are easily written for
any value of $n$, not only integer.

One should mention that, for a given lattice Hamiltonian, there are several ways to construct a CTM. The one used above is 
particularly convenient because then the RDM is just its fourth power. In general, $\rho_A$ 
will be given by the product `ABCD' of four different CTMs which also
not need to be symmetric (although the product is). 
CTMs can be defined also for non-integrable models, but then one does not have closed
and simple formulae for the eigenvalues and has to work numerically, see e.g. Ref. \cite{nishino}.

\subsection{Non-critical XXZ chain}

We start our analysis from the anisotropic Heisenberg model
with Hamiltonian
\be
H_{XXZ}=  \sum_{j}\left[ \sigma^x_j\sigma^x_{j+1}+ \sigma^y_j\sigma^y_{j+1}+\Delta
 \sigma^z_j\sigma^z_{j+1}\right] \,,
\label{HamXXZ}
\ee
where $\sigma$'s are the Pauli matrices. This model is gapless for $|\Delta|\leq1$ and gapped
for $|\Delta|>1$ with a conformal point at $\Delta=1$. We  study the
antiferromagnetic gapped regime with $\Delta>1$.
The XXZ model is usually solved by Bethe ansatz that is the most suited framework to calculate
thermodynamical properties as well as correlation functions.
Unfortunately, Bethe ansatz is still rather ineffective to provide the entanglement entropies
both in the coordinate \cite{Vidal} and algebraic \cite{afc-09} approaches.
Also the special combinatoric features at $\Delta=1/2$ only allowed the calculation of $S_n$ up to $\ell=6$
\cite{ncc-08}.
By contrast, the CTM solution is very simple and the ``single-particle levels'' of the resulting
$H_{\rm CTM}$ are exactly known \cite{pkl-99}
\be
\e_j= 2j\e  \qquad {\rm with}\qquad \e={\rm arccosh} \Delta\,.
\label{ELXXZ}
\ee
The correlation length of the model is also exactly known as function of $\Delta$ \cite{Baxter}, but for
what follows we are only interested in its universal part in the critical regime $\xi\gg1$
\be
 \ln \xi\simeq\frac{\pi^2}{2\e}+O(\e^0)\,.
 \label{xiX}
\ee

%
The R\'enyi entropies are written straightforwardly\footnote{Eq. (\ref{Sn1}) has already been derived (actually the double of it)
for the XY spin chain by Franchini et al. in \cite{fik-05}
by using previous results \cite{ijk-05} based on integrable Fredholm operators and
Riemann-Hilbert problem.
It turns out that in the XXZ model, only the dependence of $\e$ on the Hamiltonian parameters is different,
but Eq. (\ref{Sn1}) is identical. This could not have been known a-priori.
Deriving this result with CTM methods allows to generalize the treatment to other
models (as the XXZ spin-chain) for which the  Riemann-Hilbert problem is not known.
In Ref. \cite{fik-05} $S_n$ has been rewritten in terms of elliptic functions, while here we use a
different strategy.}
\be
S_n=\frac1{1-n}\left[ \sum_{j=0}^\infty\log (1+e^{-2n j\e})-n\sum_{j=0}^\infty\log (1+e^{-2j\e})\right]
\,.
\label{Sn1}
\ee
This formula is exact, but does not directly allow an expansion in terms of powers of the
correlation length, i.e. for small $\e$.
One could be tempted to perform an Euler Mac-Laurin summation, that gives the correct leading
behavior \cite{cc-04}. Unfortunately the subleading corrections we are interested in are of the
form $\xi^{-\a}\propto e^{-\a\pi^2/2\e}$, that are not  obtainable by the asymptotic expansion
in power of $\e^{-1}$ given by the Euler Mac-Laurin formula. One can however check
straightforwardly that all the analytic corrections in $\e$ indeed vanish (this has been  observed
in \cite{sce2} for $n\to\infty$).

We obtain the truly asymptotic expansion for small $\e$ by using the Poisson resummation
formula (as similarly done for $S_1$ in \cite{w-06}).
This formula tells us that
\be
\sum_{j=-\infty}^\infty f(|\e j|)=\frac2\e \sum_{k=-\infty}^\infty \hat{f} \left(\frac{2\pi k}{\e}\right)\,,
\ee
where
\be
 \hat{f}(y)=\int_0^\infty f(x) \cos (y x) d x\,.
\ee
For Eq. (\ref{Sn1}), denoting with
\be
f_n(x)= \log (1+e^{-2 n x})\,,
\ee
we can rewrite the sum as
\be
\fl S_n=\frac1{1-n}  \sum_{j=0}^\infty \left(f_n(\e j)- n f_1(\e j)\right)=
\frac12 \frac1{1-n}  \sum_{j=-\infty}^\infty \left(f_n(|\e j|)- n f_1(|\e j|)\right)+\frac{\ln2}2\,,
\ee
where we used $f_n(0)=\ln 2$.
The cosine-Fourier transform of $f_n(x)$ is
\be
\hat{f}_n (y)= \frac{n}{y^2} -\frac{\pi}{2y} {\rm csch} \frac{\pi  x}{2n}\,,
\ee
Plugging the last equation in the Poisson resummation formula we have that the two terms $n/y^2$ simplify,
leading to the compact result
\be
S_n=\frac{1}{1-n} \sum_{k=-\infty}^\infty \left[
\frac{n}{4k}  {\rm csch} \frac{ \pi ^2  k}{\e}-\frac1{4k} {\rm csch} \frac{ \pi ^2 k}{\e n}\right]
+\frac{\ln2}2\,.
\label{main}
\ee
The term $k=0$ gives the leading diverging term in the limit $\e\to 0$, and so the sum can be rewritten as
\be\fl
S_n=\frac{\pi ^2 }{24 \e}\left(1+\frac1n \right)+\frac{\ln2}2+\frac{1}{1-n} \sum_{k=1}^\infty
\left[ \frac{n}{2k}  {\rm csch} \frac{ \pi ^2  k}{\e}-\frac1{2k} {\rm csch} \frac{ \pi ^2 k}{\e n}\right]\,,
\label{sn2}
\ee
where the leftover sum over $k$ is vanishing in  the limit $\e\to 0$.
This equation is {\rm exact} for any $\e$ as it is Eq. (\ref{Sn1}) from which we started
and it is just another expansion of the elliptic function found in Ref. \cite{fik-05}.
However, in this form $S_n$ is ready for the asymptotic expansion close to $\e=0$.
For small $\e$, we have
\be
{\rm csch} \frac{\pi^2k}{\e n}\simeq 2 \exp(- \frac{\pi^2k}{\e n})\,.
\ee
Using Eq. (\ref{xiX}), we have that each of these terms gives  a correction of the form
\be
\exp(- \frac{\pi^2k}{\e n})\simeq \xi^{-2k/n}\,.
\ee
These subleading corrections clearly depend on the precise definition used for the correlation length.
Thus the universal character is only in the exponents $2k/n$, the amplitudes being
normalization dependent.
These corrections agree with the scaling form proposed in the introduction with
$x=2$. 
However, for the gapless case a multiplicative logarithmic correction is also introduced by a marginal
operator \cite{lsca-06}, but it  does not appear in the gapped phase.
This result agrees with the recent ones in Ref.  \cite{fra}.

\subsection{Other models}

Several other models can be treated with the help of CTMs
and the same procedure as above for the XXZ model applies.
This generalization is made particularly simple by the fact that in the considered cases the only change is
the expression of $\e_j$ as function of the Hamiltonian parameters.

The most studied example is the Ising model in a transverse magnetic field
with Hamiltonian
\be
H_I= - \sum_{j}\left[ \sigma^x_j\sigma^x_{j+1}+h \sigma^z_j\right] \,.
\label{HamI}
\ee
This models displays a quantum critical point at $h=1$
separating a ferromagnetic phase ($h<1$) from a quantum paramagnetic  one ($h>1$).
R\'enyi entropies for this model have been already calculated in Ref. \cite{fik-05} and so the
following results are not new, but just an asymptotic expansion of known expressions.

The ``single-particle levels'' of the resulting $H_{\rm CTM}$ are \cite{pkl-99}
\be
\e_j=
\left\{\matrix{
(2j+1)\e &\quad {\rm for} \,h>1\,,\cr
2j\e     &\quad {\rm for} \, h<1\,,
}\right.
\quad {\rm with}\;
\e=\pi \frac{K(\sqrt{1-k^2})}{K(k)}\,,
\label{EL}
\ee
where $K(k)$ is the complete elliptic integral of the first kind \cite{as},
and $k=\min [h,h^{-1}]$.
An expression for the correlation length is exactly known \cite{Baxter}, but for what follows we are
only interested in the behavior close to the critical point
\be
\ln \xi= \frac{\pi^2}{\e}+O(\e^0)\,,
\label{xiI}
\ee
valid in both phases.

In the ferromagnetic phase $h<1$, the dependence of  $\e_j$ on $j$ is the same as in the XXZ model,
only $\e$ is different. Thus Eq. (\ref{sn2}) is valid also for the Ising model.
When plugging Eq. (\ref{xiI}) into this expression, we get a leading term corresponding to the
central charge $c=1/2$ and  corrections going like $\xi^{-k/n}$, in agreement
with the presence of the energy operator of dimension $x=1$ as in the conformal case \cite{ccen-10,ce-10}.

In the paramagnetic phase $h>1$, the calculation is equivalent with the difference that
the generalized Poisson resummation formula
\be
\sum_{j=-\infty}^\infty f(|\e (b j+a)|)=\frac2{\e b} \sum_{k=-\infty}^\infty \hat{f} \left(\frac{2\pi k}{\e b}\right)
e^{2\pi i k a/b}\,,
\ee
must be used.
From Eq. (\ref{EL}), we can choose $a=1/2$ and $b=1$ (or any multiple by changing the definition of $f(x)$).
The starting formula is
\bea
S_n&=&\frac1{1-n}  \sum_{j=0}^\infty \left(f_n(\e (j+1/2))- n f_1(\e (j+1/2))\right)
\nn&=&
\frac12 \frac1{1-n}  \sum_{j=-\infty}^\infty \left(f_n(|\e (j+1/2)|)- n f_1(|\e (j+1/2)|)\right)\,,
\eea
where we notice the absence of the additive $\ln2$ term (connected with a non-degenerate ground state).
Thus, after Poisson resumming we have
\be
S_n=\frac{1}{1-n} \sum_{k=-\infty}^\infty (-1)^k \left[
\frac{n}{4k}  {\rm csch} \frac{ \pi ^2  k}{\e}-\frac1{4k} {\rm csch} \frac{ \pi ^2 k}{\e n}\right]
\,.
\label{main2}
\ee
where the main difference compared to the previous case is the factor $(-1)^k$. Thus in the expansion
in terms of the correlation length $\xi$ the same power-laws enter but with alternating amplitudes.

A particular simple further generalization is the XY chain for which all the formulas above are the same
(see e.g. \cite{p-04}) and only the expression of $k$ in terms of the $h$ and $\gamma$
changes according to
\be
k=\left\{\matrix{
\sqrt{h^2+\gamma^2-1}/\gamma    &\quad {\rm for} \,h>1\,,\cr
 \gamma/\sqrt{h^2+\gamma^2-1}   &\quad {\rm for} \, h<1\,,
}\right.
\ee
valid for $\gamma^2+h^2>1$.
This is reminiscent of the fact that at $h=1$ the transition is always in the Ising universality class and
a simple rescaling can absorb the values of $\gamma$.
Physically more interesting is the limit $h=0$, not included in the case above, but with  CTMÕs known
\cite{Baxter}. One has $\e_j = j \e$ and  $k = (1 - \gamma)/(1 + \gamma)$. The derivation of $S_n$ is
straightforward and one recovers the same corrections as for the Ising model, with a doubled leading term
reflecting the values of the central charge $c=1$.
This is connected also to an exact  relation between  Ising and XX models \cite{ij-08}.

Also XYZ chains (see \cite{eer-09} for $S_1$)
and spin $\kappa/2$ analogue of the XXZ quantum spin chain (see \cite{w-06} for $S_1$) can be
simply obtained from the formulas above.

\subsection{The single-copy entanglement}\label{singlecopy}

The limit $n\to\infty$ requires a separated analysis, because
in Eq. (\ref{Sgapcorr}) the exponent of the corrections goes to zero, signaling the appearance of
logarithmic corrections to the scaling as for the conformal case.
$S_\infty$ is  called single copy entanglement \cite{sce,sce3,sce2} and it
is obtained by taking the limit $n\to\infty$ before the limit $\e\to0$, since the two do not commute.
It is easy to get the logarithmic corrections exactly. For $n\to\infty$, Eq. (\ref{sn2}) becomes
\bea
S_\infty&=&\frac{\pi ^2 }{24 \e}+\frac{\ln2}2 +\sum_{k=1}^\infty
\left[ \frac{\e}{2\pi^2 k^2} - \frac{1}{2k}  {\rm csch} \frac{ \pi ^2  k}{\e}\right]\\
&=&\frac{\pi ^2 }{24 \e}+\frac{\ln2}2 +
 \frac{\e}{12} - \sum_{k=1}^\infty \frac{1}{2k}  {\rm csch} \frac{ \pi ^2  k}{\e}
 \label{Sinf2}
\,. \eea The left over sum gives standard power-law corrections
and they do not involve `unusual' exponents. The unusual part is
in the logarithmic corrections encoded in the term $\e/12$. Indeed
using $\ln \xi=\pi^2/(a \e)$ (with $a=1$ for Ising ferromagnetic
and $a=2$ for XXZ), we have
\begin{equation}\label{singlecorr}
S_\infty-S_\infty^{\rm asy}\simeq \frac{\pi^2}{12 a\log \xi}.
\end{equation}
Notice that when written in terms of $\e$ this expression has a unique correction, that is
particularly simple for the XXZ chain since $\e={\rm arccosh} \Delta$.

For the Ising model in the paramagnetic phase $h>1$, the terms $(-1)^k$ change the $1/\e$ term,
and the final result is
\be
S_\infty=\frac{\pi ^2 }{24 \e}-
 \frac{\e}{24} - \sum_{k=1}^\infty \frac{(-1)^k}{2k}  {\rm csch} \frac{ \pi ^2  k}{\e}
\,.
\label{Sinf4}
\ee
Notice that the amplitudes of the corrections are different in the two phases and their numerical
values are of the same form of what found for the critical XX spin-chain \cite{ce-10}.
These results for the single copy entanglement in XY chains have been reported also in \cite{sce2,sce3}.

\section{Relation to Virasoro characters}

The above examples may be considered as special cases of a more
general phenomenon: that in many integrable models satisfying the
Yang-Baxter relations, traces of powers of the CTM may be
expressed in terms of the Virasoro characters of the corresponding
CFT of the critical theory. These enjoy simple properties under
modular transformations which allow us to extract the `unusual'
corrections to scaling discussed earlier.

To be specific, the Boltzmann weights of many integrable models
lie on a curve specified by an elliptic modulus $q$, where the
critical point corresponds to $q\to1$. 
The examples of quantum spin chains
considered above correspond to 
$q=e^{-2\epsilon}$. Baxter \cite{Baxter} showed that in these
models the eigenvalues of the 4th power ${\hat A}^4$ of the CTM
are all of the form $q^{aN+b}$, for integer $N\geq0$, with a
computable degeneracy $d_N$. It was then observed in the late
1980s (see, for example, Refs.~\cite{JCVir, JClh}) that the
degeneracy factors are just those appearing in the highest weight
representations of the Virasoro algebras in the CFT which
describes the scaling limit at the critical point, that is
\begin{equation}\label{eq:chi}
\chi_{\Delta}(q)=q^{-c/24+\Delta}\sum_{N=0}^\infty d_Nq^N\,,
\end{equation}
where $c$ is the central charge and $\Delta$ is the highest weight
of the representation, giving the dimension $x=2\Delta$ of a
scaling operator of the theory. The value of $\Delta$ depends on
the value of the spin at the origin, and the boundary conditions
at infinity. This observation, although not proven in general, was
based on the facts that (a) in the scaling limit at the critical
point, but in a finite annulus with outer and inner radii $R_>$
and $R_<$, the eigenvalues of the CTM have precisely this form,
with conjugate modulus $\tilde q$ (defined below) equal to
$R_</R_>$ -- this follows from boundary CFT \cite{JCBCFT}; and (b)
in the integrable RSOS models, the eigenstates of the CTM
correspond to a semi-infinite walk on the corresponding $A_m$
diagram, and the Virasoro generators act simply on the space of
these paths. (In fact it was shown by the Kyoto group \cite{Kyoto}
that for higher integrable models obtained by fusion the result is
not always a simple character but rather a branching coefficient
of an affine algebra. However these still enjoy simple modular
properties as described below.)

This means, in particular, that the partition function ${\rm
Tr}\,{\hat A}^4$ is proportional to $\chi_\Delta(q^a)$ for some
$\Delta$ (which depends on, for example, which massive phase the
model is in). Without loss of generality we take $a=1$. In general
the partition function may be a linear combination of characters,
depending on the choice of boundary conditions at the origin and
infinity. This complicates the argument without changing the
general conclusion, and we shall assume that the boundary
conditions are such as to pick out a single character, as in the
examples discussed earlier. However we remark that, in the context
of entanglement entropy, we should specify if we fix the spin at
the origin in the direct or in the dual lattice. In all examples
above, we fixed the dual variable, by leaving the bond between the
two halves free. However, while at first it could sound strange,
also fixing an actual spin at the origin makes sense for the
entanglement entropy, because in the CTM this does not correspond
to fixing it in the hamiltonian (which would effectively divide
the chain into two non-interacting halves). Rather it means
projecting the ground state $|0\rangle$ into a subspace in which
the spin at the origin is fixed, and measuring the entanglement
between the two halves of the chain in this subspace. Thus we
expect to be able to explore other values of $\Delta$ by such a
procedure, and will therefore keep it general in what follows.

The point now is that the trace of ${\hat A}^{4n}$ is, in the
basis in which the CTM is diagonal, given simply by
replacing $q\to q^n$, and thus
\begin{equation}\label{rhoAchar}
{\rm Tr}\,\rho_{\cal
A}^n=\frac{\chi_\Delta(q^n)}{\big(\chi_\Delta(q)\big)^n}\,.
\end{equation}

We are interested in the limit $q\to1$ close to the critical
point, where the series expressions (\ref{eq:chi}) for
$\chi_\Delta(q)$ are not very useful. However \cite{JCmod}, the
Virasoro characters transform linearly under a modular
transformation $q\to\tilde q$, where, if $q=e^{-2\epsilon}$,
$\tilde q=e^{-2\pi^2/\epsilon}$:
\begin{equation}
\chi_{\Delta}(q)=\sum_{\Delta'}S_\Delta^{\Delta'}\chi_{\Delta'}(\tilde
q)\,,
\end{equation}
where $S_\Delta^{\Delta'}$ is the modular S-matrix which
characterizes the CFT. Notice that the modular invariance
plays exactly the same rule as Poisson resummation above, but
it is more general.

As $q\to1$, $\tilde q\to0$, so it is straightforward to extract
the leading behavior and all the corrections. The leading term
comes from $\Delta'=0$, in which case we see that
$$
{\Tr}\,\rho_{\cal A}^n\sim \big(S_\Delta^0\big)^{1-n}\left({\tilde
q}^{-c/24}\right)^{n^{-1}-n}\,,
$$
which gives
\begin{equation}\label{eqSnmod}
S_n\sim -\frac c{24}\left(1+\frac1n\right)\ln\tilde
q+\ln\big(S_\Delta^0\big)+o(1)\,.
\end{equation}
Comparing with (\ref{Renyi:asymp-gap}), we see that $\tilde
q\propto\xi^{-2}$. The $O(1)$ term is related to the
Affleck-Ludwig boundary entropy \cite{AffleckLudwig}
$g_\Delta=\ln\big(S_\Delta^0\big)^{1/2}$. Note that we get twice
the boundary entropy because there are two semi-infinite chains
adjoining the spin at the origin. The examples discussed earlier
presumably correspond to $\Delta=0$ and the corresponding $O(1)$
term can be absorbed into the non-universal constant of
proportionality between $\tilde q$ and $\xi^{-2}$.

The corrections now come from (a) integer powers of $\tilde q$,
that is powers of $\xi^{-2}$, in the expansion of the character,
and (b) other characters with $\Delta'>0$. The leading terms of
the latter form then give rise to corrections to (\ref{eqSnmod})
\be
(1-n)^{-1}\sum_{\Delta'>0}\frac{S_\Delta^{\Delta'}}{S_\Delta^0}\left({\tilde
q}^{\Delta'/n}-n{\tilde q}^{\Delta'}\right)+\cdots\,.
\ee
(The second power is less important for $n>1$ but is included so
as to ensure the finiteness as $n\to1$.) Recalling that
$2\Delta'=x'$, we see that we get corrections of the form
$\xi^{-x'/n}$ as expected. However, unless the matrix element
$S_{\Delta}^{\Delta'}$ happens to vanish, in principle we get such
contributions from \em all \em the scaling dimensions $x'$ of
primary fields, not just the relevant ones. Also, the higher order
corrections in general involve powers which are all possible
integer linear combinations of $x'/n$ and $x'$.
The evidence of these further corrections has been reported for
simple gapless spin chains \cite{ce-10}, but it would be as
interesting to report them in the gapped case.

We can also derive a general result for the single-copy
entanglement $S_\infty$. Using 
Eq. (\ref{rhoAchar}) we see that
\be
S_\infty=-(-c/24+\Delta)\ln q+\ln\chi_\Delta(q)\,.
\label{Sinf3}
\ee
The second term, after using the modular transformation, goes like
$(-c/24+\Delta')\log\tilde q$ plus power-law corrections. 
We expect the leading term in the sum over $\Delta'$ to come from
$\Delta'=0$. Thus we get a universal term $(c/12)\ln\xi$. This
corresponds to the first term in Eqs. (\ref{Sinf2}) and  (\ref{Sinf4}).
The first term in Eq. (\ref{Sinf3}) gives the `unusual' correction discussed in
Sec.~\ref{singlecopy}. Since $\tilde q\propto\xi^{-2}$, $\ln
q\sim-2\pi^2/\ln\xi$, so this correction has the form
\be
\frac{2\pi^2(-c/24+\Delta)}{\ln\xi}\,.
\ee
This agrees with  the Ising case in the  ferromagnetic phase given by Eq.  (\ref{singlecorr}) using
$c=1/2$, $\Delta=1/{16}$, and in the paramagnetic Eq. (\ref{Sinf4}) using $c=1/2$ and $\Delta=0$.

\section{Concluding remarks}

We showed that the corrections to the scaling of the R\'enyi entanglement entropies in
gapped systems display the universal form (\ref{Sgapcorr}) in the case when an infinite line is
divided into two semi-infinite subsystems.
We provided few explicit examples using known corner transfer matrices and we argue that
in the general case these corrections are a consequence of the modular invariance of the
traces of powers of the CTM when expressed in terms of the Virasoro characters.

This result generalizes straightforwardly to more complicated bipartitions provided the correlation
length $\xi$ is smaller than all separations. Eq. (\ref{Sgapcorr}) only gets multiplied
by the number of boundary points between ${\cal A}$ and ${\cal B}$.
The rigorous result for a finite interval of length $\ell$ obtained for the XY model \cite{fik-05} confirms this.
However, for a finite interval, only  when $\ell$ becomes much larger than $\xi$,
the asymptotic behavior would be visible. For smaller values of $\ell$, a complicated crossover
between the conformal result and the asymptotic one takes place as already known for the
leading terms \cite{cd}.
The characterization of all these corrections is not only an academic task.
In the case of the entanglement of two disjoint intervals in a conformal model,
their precise knowledge has been fundamental
to recover the CFT predictions \cite{fps-09,cct-09} in numerical calculations \cite{cg-08,atc-09,fc-10,ip-09}.
We believe that the same is true for massive systems, especially in the case of non-integrable models
when numerical calculations are the only way to attack the problem.
A final question is whether these unusual corrections are also present for other entanglement measures
that displayed them in the gapless phases as the valence bond entanglement  \cite{ar-10},
and if yes whether it is possible to calculate them exactly.

\section*{Acknowledgments}

This work has been done in part at the Max Planck Institute for
the Physics of Complex Systems in Dresden whose hospitality is
kindly acknowledged. We thank Masud Haque for fruitful
discussions. PC is highly indebted to Fabian Essler for the
collaboration in Ref. \cite{ce-10} that shares many analogies
with the problem studied here. 
We are grateful to all the authors of Ref [22] for signaling us a major typo in the first version of this paper.
JC was supported in part by EPSRC Grant
EP/D050952/1.

\end{document}